\documentclass{npw}

\usepackage{graphicx}
\usepackage{amssymb}
\def\eV{\hbox{ eV}}
\def\Tr{\hbox{ Tr}}
\def\Prob{\hbox{ Prob}}
\def\rect{\hbox{ rect}}
\def\EOp{\mathsf{E}\kern-1pt\llap{$\vert$}}

\begin{document}

\title{On particle oscillations}

\author{
  Marek G\'o\'zd\'z\email{mgozdz@kft.umcs.lublin.pl} \\
  {\it Department of Informatics, Maria Curie-Sk{\l}odowska University} \\ 
  {\it ul. Akademicka 9, 20-033 Lublin, Poland} \\
  Andrzej G\'o\'zd\'z \\
  {\it Department of Physics, Maria Curie-Sk{\l}odowska University} \\
  {\it pl. Marii Curie--Sk{\l}odowskiej 1, 20-031 Lublin, Poland} 
}


\date{\today}

\maketitle

\begin{abstract}
  It has been firmly established, that neutrinos change their flavour
  during propagation. This feature is attributed to the fact, that each
  flavour eigenstate is a~superposition of three mass eigenstates, which
  propagate with different frequencies. This picture, although widely
  accepted, is wrong in the simplest approach and requires quite
  sophisticated treatment based on the wave-packet description within
  quantum field theory. In this communication we present a~novel, much
  simpler explanation and show, that oscillations among massive
  particles can be obtained in a~natural way. We use the framework of
  quantum mechanics with time being a~physical observable, not just
  a~parameter.
\end{abstract}

\section{Introduction}

All elementary particles are subject to mixing within their respective
groups, \textit{i.e.}, quarks, neutral leptons (neutrinos), charged
leptons, as well as gauge bosons. This peculiar feature of gauge
theories underlying the Standard Model comes from the requirement that
the quantum numbers should match those observed in nature. In other
words, in order to arrive at a~picture consistent with the experiment,
one has to `rotate' sectors of the Standard Model, with the rotation
parameters fitted from the experimental data. As a~consequence of
mixing, particles should oscillate between their possible states, as it
is observed for neutrinos and some mesons.

However, from the theoretical point of view, not everything is clear in
this picture. Take neutrino oscillations as an example. The widely
accepted explanation is based on the assumption, that neutrinos which
are paired with the charged leptons ($e$, $\mu$, $\tau$) are not the
same as the propagating neutrinos. From the so-called interaction basis
the interaction eigenstates $\nu_\alpha$, $\alpha=e,\mu,\tau$ have to be
rotated by a~unitary matrix to the physical basis, in which the states
$\nu_i$, $i=1,2,3$ are those which propagate. The latter are physical
particles with well defined masses, while the former are ill-defined,
therefore virtual particles.

The immediate question which arises is, why do we have to work in two
bases -- one to describe interactions (with unrealistic particles with
no definite mass) and another to describe propagation (of physical
particles which are not observed in nature as standalone objects)? This
counter-intuitive picture poses even more trouble when one wants to
formulate a~consistent description of, say, neutrino oscillations. Even
assuming that in the process of emission three different physical
particles are produced, each with well defined mass, momentum, and
energy, it is difficult to justify, how these particles can arrive at
the detection point as a~single, detectable object. To resolve this
problem different authors have argued, that these particles should share
a~common momentum or a~common energy. Curiously, both approaches lead to
the same final expression for the phase of oscillations. The same
expression can also be reached when assuming nothing but neutrinos being
ultra-relativistic \cite{nu-paradox}. The most correct derivation of the
neutrino oscillations phase involves a~full wave-packet treatment within
quantum field theory \cite{beuthe}.

In this communication we propose a~different mechanism which leads to
particle oscillations. Without referring to two different classes of
states and working with the physical particles only we show, that under
certain assumptions transitions between mass eigenstates can be
observed. Our framework is the quantum mechanics in which time is no
longer a~parameter but one of the space-time variables.

\section{The model}

Recent experimental progress in the field of quantum mechanics suggests,
that the ordinary formulation is not enough to properly describe what is
being observed. In the so-called delayed-choice experiments
\cite{Wheeler78,Grang86,Jacques07} the cause and consequence seem to be
inverted in time, implying that either causality is violated or our
understanding of quantum phenomena should be altered. Also the newest
experiments involving entangled systems \cite{qm-ent1} led to the
conclusion, that within the traditional framework of quantum mechanics
and special relativity, superluminal communication between different
parts of the system is observed unless we change some basic principles
in the formalism. Only recently, an entangled system of two photons that
never co-existed in time has been created \cite{qm-ent2}. Another
example is the observation of interference fringes \cite{qm-int1} which
are in agreement with the hypothesis, that the wave functions interfere
in time, not in spatial variables.

Motivated by this line of research, a~new quantum theory seems
desirable, and one of such models has been proposed in
Ref.~\cite{ag-qm}. One of its main features is the inclusion of time as
an observable, such that it is possible to consistently construct a~time
operator \cite{ag-timeOp}. Consequently, no time evolution of the wave
function, which is space as well as time dependent, is needed. Each
measurement is represented by a~projection of the wave function on the
states of a~properly constructed `detector', according to the Dirac
projection postulate. This model successfully described such phenomena
as: arrival time \cite{aq-ArrivT}, delayed choice in quantum mechanics
\cite{ag-qm} and interference in time \cite{qm-int2,qm-int3}.

In this communication we outline the description of the oscillations of
mass eigenstates, which ultimately can be used to describe neutrino
oscillations.

\section{Particle oscillations}

Keeping in mind neutrinos as our primary example, we want to show, that
after emitting a~particle of certain mass, another particle of a~close
laying mass can be observed. This all can happen under the assumption,
that these particles share most if not all other properties,
\textit{i.e.}, spin, electric charge etc.

We divide the process of describing particle oscillations into three
stages:
\begin{enumerate}
\item the emission (creation) of the particle denoted by $\tau_1$,
\item propagation of the particle; for the sake of simplicity we assume
  free propagation and denote this stage by $\tau_2$,
\item detection of the particle ($\tau_3$).
\end{enumerate}

According to the general rules given above, one has to construct
projection operators describing each stage, and project the initial wave
function of the emitted particle subsequently using the appropriate
operators. The final outcome will represent the probability density of
detecting the particle. Denoting by $\psi$ the initial wave function,
the density matrix after the first stage is given by
\begin{equation}
  \label{eq:st1}
  \rho_1(\tau_1) = |\psi\rangle\langle\psi|.
\end{equation}
At the second stage the state changes into $\rho_2$ given by
\begin{equation}
  \label{eq:st2}
  \rho_2(\tau_2) = \frac{\EOp(\tau_2) \rho_1(\tau_1) \EOp(\tau_2)}
  {\Tr[\EOp(\tau_2) \rho_1(\tau_1) \EOp(\tau_2)]},
\end{equation}
where Tr denotes the trace, which provides proper normalisation of the
expression, $\EOp(\tau_2)$ is the projection operator which here
describes the propagation, and $\rho(\tau_1)$ is given by
Eq.~(\ref{eq:st1}). The detection process introduces yet another
operator $\EOp(\tau_3)$ which defines the detector and acts according to
\begin{equation}
  \label{eq:st3}
  \rho_3(\tau_3) = \frac{\EOp(\tau_3) \rho_2(\tau_2) \EOp(\tau_3)}
  {\Tr[\EOp(\tau_3) \rho_2(\tau_2) \EOp(\tau_3)]}.
\end{equation}
Now, the probability of the process $\tau_1\to\tau_2\to\tau_3$ is given
by
\begin{equation}
  \label{eq:rho2}
  \Prob(\tau_1\to\tau_2\to\tau_3) = 
  \Tr[\EOp(\tau_3)\EOp(\tau_2) \rho_1(\tau_1) \EOp(\tau_2)\EOp(\tau_3)].
\end{equation}

To be more specific, let us assume three members of a~family of
particles having very similar masses. In the case of neutrinos, the mass
differences are of the order of 10 meV or less which indicates really
close laying states. Assume further that one of these particles is being
created in a~reaction. In the usual approach one neglects the time this
reaction takes. It implies further that the produced particle appears
immediately in zero-time, with sharp defined mass. On the other hand, if
one takes into account that the reaction time is non-zero and finite,
this introduces a~kind of uncertainty in time for the particle to be
produced, which results in a~broadening of its energy profile. Thus, the
created particle is no longer sharply peaked in mass, but possesses also
an uncertainty in this parameter. One may also justify the broadening of
mass of a~particle in a~more formal way. Namely, in our model the mass
operator does not commute with the time operator. Therefore a~kind of
uncertainty relation between mass and time can formally be given, which
prevents sharply peaked mass distribution to appear in any finite time.

\begin{figure}
  \centering
  \includegraphics[width=0.85\columnwidth]{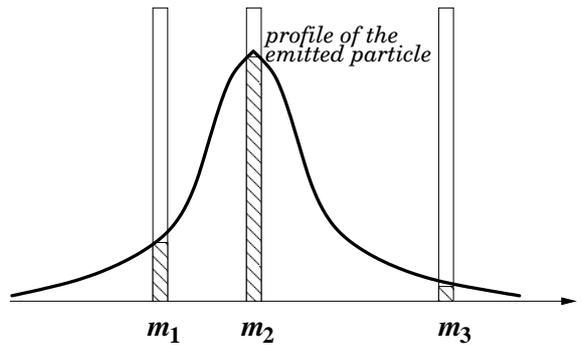}
  \caption{Schematic picture of the mass profile of the emitted
    particle. The emitted particle of central mass $m_2$ effectively has
    admixtures from masses $m_1$ and $m_3$. }
  \label{fig:profile}
\end{figure}

The broadened mass profile overlaps with the neighbouring mass states,
effectively turning into a~linear combination of states with different
masses, with the `mixing parameters' given by the values of the function
describing the profile (see Fig.~\ref{fig:profile} for a~graphical
representation).

Let us work, without loss of generality, in two dimensions, one time and
one spatial variable $(t,x)$. It follows, that the wave function of the
emitted particle may be written in the form
\begin{equation}
  \label{eq:psi}
  \psi(\tau_1;t,x) = 
  \int_{\mathbb{R}^2} d^2k \ a_M(k_0,k_1) \ \eta^*_k(t,x),
\end{equation}
where $\tau_1$ denotes the first stage of the process, $\eta^*_k(t,x) =
\eta^*_{k_0}(t)\ \eta^*_{k_1}(x) = \exp(-\mathrm{i}k_0 t)
\exp(-\mathrm{i}k_1 x)/2\pi$, and the function $a_M(k_0,k_1)$ of the
2-momentum describes the shape of the particle in the momentum space. We
notice for future reference, that $k_0^2-k_1^2=m^2$, so that it is
possible to change the integration variables from $(k_0, k_1)$ to $(m^2,
k_1)$.\footnote{We use natural units: $\hbar=c=1$.}

The rules of the free propagation of the particle are governed by the
structure of the vacuum. From this point of view the vacuum cannot
distinguish the broadened state $\psi(\tau_1)$ from three separate mass
states propagating together. We therefore construct the projection
operator describing the propagation as
\begin{equation}
  \label{eq:E}
  \EOp(\tau_2) \psi(\tau_1;t,x) = 
  \int_{\Delta_k} d^2k \ \langle\eta_k^*(t,x)|
  \psi(\tau_1;t,x)\rangle \eta_k^*(t,x),
\end{equation}
where $\Delta_k$ is the set of 2-momenta, that can be transmitted
through the vacuum during propagation. This, after a variable change
$(k_0,k_1) \to (m^2,k_1)$, using the mass-shell relation, turns into
a~disjoint set of narrow peaks around $m_1$, $m_2$, and $m_3$. We
therefore assume, that the structure of the vacuum permits propagation
of some chosen set of masses, which defines our Standard
Model. Evaluating Eq.~(\ref{eq:E}) using (\ref{eq:psi}) one gets
\begin{equation}
  \label{eq:Epsi}
  \EOp(\tau_2) \psi(\tau_1;t,x) = 
  \int_{\Delta_k} d^2k \ a_M(k_0,k_1) \eta_k^*(t,x).
\end{equation}

Finally, let us denote the wave function of the detector by
$\phi(\tau_3;s,X)$. The detector is two-dimensional (one time and one
spatial dimension) and located at the space-time point $(s,X)$. Notice,
that by construction the measurement is time dependent. The projection
operator representing the detection is now
\begin{equation}
  \label{eq:E3}
  \EOp(\tau_3) =  |\phi(\tau_3;s,X)\rangle\langle\phi(\tau_3;s,X)|.
\end{equation}
If we want to distinguish the detected particles by their masses, the
detector function $\phi$ should describe states with definite mass, or
should be peaked around some mass in the mass-momentum space. One such
example is the infinite potential well described in the next section.

\subsection{Example: infinite potential well}

Let us represent the detector as eigenfunctions of the Klein--Gordon
equation in a~two-dimensional infinite potential well
$\phi^{(s,X)}_{n_0,n_1}(\tau_3;t,x)$. Denote its dimensions and
localisation by $L_0 \times L_1$ with the central point $(s+L_0/2,
X+L_1/2)$, \textit{i.e.}, it is a~rectangle with the closer corner given
by the space-time point $(s,X)$, extending by $L_0$ in the time
direction and by $L_1$ in the spatial direction. The detector has to be
tuned to detect certain mass $m_D$ given by
\begin{equation}
  (m_D)^2 = \pi^2 \left( \frac{n_0^2}{L_0^2} -
    \frac{n_1^2}{L_1^2} \right)
\end{equation}
where $n_0$, $n_1$ sign different modes of the wave function within the
well. The probability of detection is in this case given by
\begin{eqnarray} 
  && \Prob(\tau_1\to\tau_3;s,X) =  \\
  && \sum_{n_0,n_1} \left | \int_{\Delta_k} d^2k \ 
    \langle\eta_k^*| \psi(\tau_1;x) \rangle
    \langle\phi^{(s,X)}_{n0,n_1}(\tau_3;t,x) | \eta_k^* \rangle 
  \right |^2.  \nonumber
\end{eqnarray}
Working out this example explicitly, the full formula reads
\begin{eqnarray} 
  && \Prob(\tau_1\to\tau_3;s,X) =  \nonumber \\ 
  && {\cal N} \sum_{n_0,n_1} (n_0 n_1)^2 \Bigg |
  \int_{\Delta_{(m^2)}} d(m^2) \int_\mathbb{R} dk \
  \frac{a_M(m^2,k)}{\sqrt{m^2+k^2}} 
  \nonumber \\
  &&\times \exp\left\{ -{\rm i} \left[
      \sqrt{m^2+k^2} L_0(s+1) + k (X+L_1)\right] \right\} 
  \nonumber \\
  &&\times \frac{\sin \left(\sqrt{m^2+k^2}\frac{L_0}{2} -
      n_0\frac{\pi}{2}\right)} {(m^2+k^2)L_0^2 - (n_0 \pi)^2} \
  \frac{\sin \left(k\frac{L_1}{2} - n_1\frac{\pi}{2}\right)}{(k L_1)^2 -
    (n_1 \pi)^2} \Bigg |^2,
  \nonumber \\
\end{eqnarray}
where we have changed the variables to integrate over masses
squared. Here $\cal N$ is an overall normalisation factor, and we recall
that in normal units all masses should be read as $m/(\hbar c)$.

The region $\Delta_{(m^2)}$ consists of three narrow peaks around the
masses $m_1$, $m_2$, and $m_3$. One may therefore simplify the integral
over $m^2$ and substitute it by a sum
$$
\int_{\Delta_{(m^2)}} d(m^2) F(m^2,k) \to \sum_{j=1,2,3} (4 m_j
\delta_2) F(m_j^2,k) 
$$
where $\delta_2$ is the (common, for simplicity) width of the peaks
around the masses $m_j$, characteristic for the second stage
$\tau_2$. This shows clearly that in the final formula (after the
modulus squared is applied), interference terms involving different
masses $m_j$ will appear, leading to possible oscillations.  We have
shown therefore, that the emission of one mass results in our model in
a~non-zero probability of detection of another mass. This probability is
some function of the localisation of the detector in space-time. The
detailed behaviour depends strongly on the construction of the initial
wave function ($\psi(\tau_1)$) and the detector ($\phi(\tau_3)$). We
will discuss a~simpler numerical example in the next section.

\section{The truncated cosine distribution}

\begin{figure}
  \centering
  \includegraphics[clip,width=0.7\columnwidth]{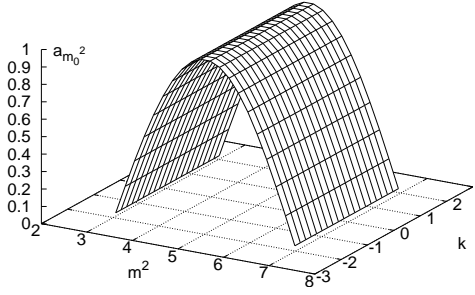}
  \caption{Plot of the profile function $a_{m_0^2}(m^2,k)$
    Eq.~(\ref{eq:aMm}) for $m_0^2=5$, $\delta=2$, $W=2$.}
  \label{fig:aM}
\end{figure}

To better control which mass is emitted, let us write down the initial
profile $a_M(k_0,k_1)$ as $a_{m_0^2}(m^2,k_1)$, with $m_0$ being the
central value of the emitted mass. We propose in this example the
following explicit form of the profile function:
\begin{eqnarray}
  \label{eq:aMm}
  a_{m_0^2}(m^2,k_1) &=& \cos \left((m^2-m_0^2)\frac{\pi}{\delta} \right)
  \rect_\delta(m^2-m_0^2) \nonumber \\
  &\times& \rect_W(k_1),
\end{eqnarray}
which in the momentum space is given by
\begin{eqnarray}
  \label{eq:aMk}
  a_{m_0^2}(k_0,k_1) &=& 
  \cos \left((k_0^2-k_1^2-m_0^2)\frac{\pi}{\delta} \right) \\
  &\times& \rect_\delta(k_0^2-k_1^2-m_0^2) \rect_W(k_1). \nonumber
\end{eqnarray}
Here the box function $\rect_w(x)$ is a~rectangle of width $w$, height
1, centred around 0 in the $x$ variable. The distribution
(\ref{eq:aMm}) gives a~cosine-shaped smearing of width $\delta$ in the
masses squared around $m_0^2$, and a~flat smearing in spatial momenta of
width $W$. This choice is physically reasonable. The function
$a_{m_0^2}(m^2,k_1)$ is depicted on Fig.~\ref{fig:aM} for some choice of
the width parameters.

We further assume, that the process of detection is in fact quite
similar to the process of emission. In the case of neutrinos we expect
them to be created and detected in a~weak process, so the interaction
vertexes in the Feynman diagrams are similar (\textit{e.g.} a~beta and an
inverse-beta decay). Therefore we define both the initial wave function
and the detector wave function in a~similar way, \textit{i.e.},
\begin{eqnarray}
  && \psi(t,x) = \int_{\mathbb{R}^2} d^2k \ a_{m_0^2}(k) \ \eta^*_k(t,x), \\
  && \phi(t,x) = \int_{\mathbb{R}^2} d^2k \ a_{m_D^2}(k) \ \eta^*_k(t,x),
\end{eqnarray}
$m_0$ and $m_D$ being the emitted and detected mass, respectively. Now,
the detector needs to be shifted from the origin to the point $(s,X)$
resulting in
\begin{equation}
  \phi(t,x) \to \phi(t-s,x-X) = 
  e^{\mathrm{i} k_0 s} e^{\mathrm{i} k_1 X} \phi(t,x).
\end{equation}
Finally, we arrive at the following formula for the probability of
detection:
\begin{eqnarray}
  && \Prob(s,X) = {\cal N} \Bigg |
  \int_{\mathbb R} dk_1 
  \int_{\Delta_{(m^2)}} d(m^2) \nonumber \\ 
  && \frac{e^{\mathrm{i} \sqrt{m^2+k_1^2} s} e^{\mathrm{i} k_1 X}} 
  {\sqrt{m^2+k_1^2}}
  a_{m_0^2}(m^2,k_1) a_{m_D^2}(m^2,k_1)
  \Bigg |^2, 
\end{eqnarray}
where $\cal N$ is an overall normalisation factor and the functions $a$
are given by Eq.~(\ref{eq:aMm}). The integration range $\Delta_{(m^2)}$
consists of three narrow peaks around the masses $m_{1,2,3}$. Assuming
the width of the peaks $\delta_2$ being small, one may approximate this
integration by taking the value of the integrand in the central points
times the width of the peaks, which yields
\begin{eqnarray}
  \label{eq:Prob}
  && \Prob(s,X) = {\cal N} \Bigg |
  \int_{\mathbb R} dk_1 \sum_{j=1,2,3} (4 m_j \delta_2)
  \nonumber \\ 
  && \frac{e^{\mathrm{i} \sqrt{m^2+k_1^2} s} e^{\mathrm{i} k_1 X}}
  {\sqrt{m_j^2+k_1^2}}
  a_{m_0^2}(m_j^2,k_1) a_{m_D^2}(m_j^2,k_1)
  \Bigg |^2.
\end{eqnarray}
Notice that both $m_0$ and $m_D$ are one of the $m_j$'s.

\begin{figure}
  \centering
  \includegraphics[width=0.7\columnwidth]{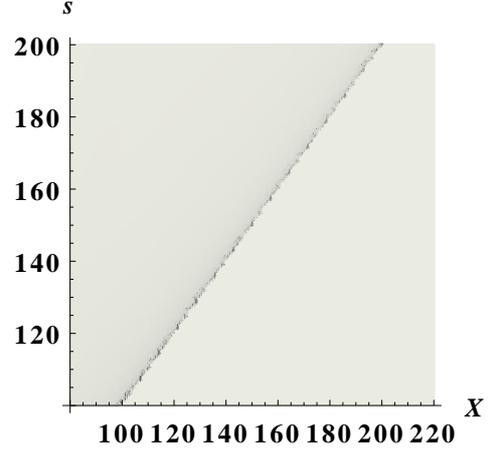}
  \caption{Density plot of the probability $\Prob(s,X)$ for $m_0=m_3$
    and $m_D=m_1$. A~clear maximum is visible for $s=X$.}
  \label{fig:3D}
\end{figure}

First of all let us check the overall behaviour of the formula
Eq.~(\ref{eq:Prob}). We present a~3D plot of the values of Prob as
a~function of the localisation $(s,X)$ of the detector on
Fig.~\ref{fig:3D}. This is a~neutrino-inspired example with the masses
given by \cite{pdg2012}
\begin{eqnarray*}
  \label{eq:params}
  & m_1 = 0.1\eV, 
  \quad m_2=m_1+\Delta m^2_{12}, \quad m_3=m_1+\Delta m^2_{13} &  \\
  & \Delta m^2_{12} = 7.6\times10^{-5} \eV^2, \qquad
  \Delta m^2_{13} = 2.5\times10^{-3} \eV^2. &
\end{eqnarray*}
The smearing parameters where chosen as
$$
  \delta_0 = \delta_D = 1.1 \times \Delta m^2_{13}, \qquad
  \delta_2 = 0.001 \times \Delta m^2_{12}.
$$

\begin{figure}[t]
  \centering
  \includegraphics[width=0.7\columnwidth]{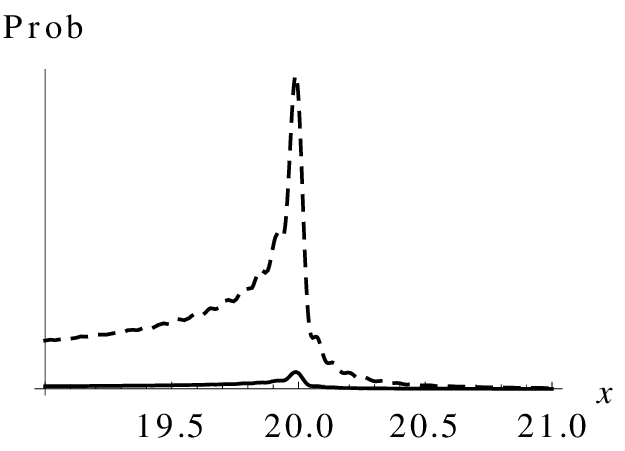} \\
  \includegraphics[width=0.7\columnwidth]{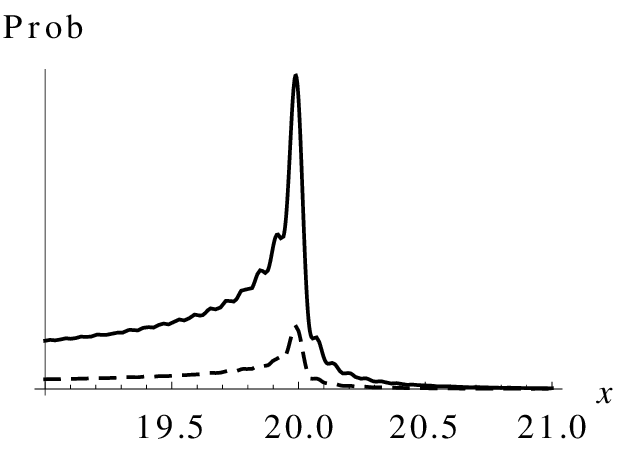}
  \caption{Not normalised probabilities Eq.~(\ref{eq:Prob}) of detection
    of $m_1$ (dashed lines) and $m_3$ (solid lines) if the mass
    $m_0=m_1$ (upper plot) and $m_0=m_3$ (lower plot) is emitted.}
  \label{fig:probD}
\end{figure}

We first notice, that there is a~strong maximum of the detecting
probability along the $s=X$ line. This indicates, that the particles
propagate with some maximum speed which corresponds approximately to the
speed of light (in our units $c=1$). So the choice of small masses imply
the particles being ultra-relativistic. Another feature is, that the
probability is exactly zero until the fastest particles reach the point
$s=X$ (lower light triangle), but remains non-zero for later times, as
slower particles (or tails of the wave functions of particles that have
already passed this point) may still be detected. This presents
a~physically consistent picture of particle propagation. Notice, that on
Fig.~\ref{fig:3D} the case of $m_0\not=m_D$ is demonstrated.

A~detailed analysis is shown on Fig.~\ref{fig:probD}, on which we
present the shapes of the detection probability functions for time
$s=20$. We include the probabilities of detection of $m_1$ and $m_3$,
when $m_1$ or $m_3$ is emitted. The curve describing $m_2$ will be lying
in between these two. The probability is not normalised, so the units on
the vertical axes are arbitrary.  One clearly sees, that the probability
shape is strongly peaked around $x=20$, which corresponds to the choice
of $s$, and vanishes to zero for greater distances. Also, the maximum
probability is obtained for the emitted mass, but some admixture of the
other mass is also observed. This will lead to a~drop in the observed
flux of the particles, which is widely regarded as the proof of particle
undergoing oscillations. In fact, with our choice of the profile
function $a_{m^2}$ being proportional to the cosine, some oscillations
of the probability of detection are also visible. These oscillations are
better visible after the inclusion of proper normalisation ${\cal N}$,
but we will discuss this topic in more detail in an upcoming paper.

\section{Conclusions and outlook}

We have shown, that particle oscillations may be a~pure quantum
mechanical phenomenon and do not require invoking unphysical
interaction eigenstates. In our model the effect comes from the
uncertainty principle after taking into account the fact, that no
physical process can happen in zero-time. The broadening of mass
spectrum of the emitted particle implies overlaps with the neighbouring
mass states, which effectively adds them as admixtures to the
propagating state. In this model it is natural, that heavier particles
(like charged leptons for example) will not mix due to huge mass
differences between them. The mixing of light quarks and almost no
mixing between charged leptons is in excellent agreement with
observations.

More work is needed to check, to which extent the effect depends on the
detailed choice of the $a_{m^2}$ profile. We suspect, that each
distribution (like Gaussian, inverted parabola etc.) should lead to
similar results, but no formal proof of this statement has been
constructed. It is also interesting to investigate this problem for
different constructions of the detector.

\begin{ack}
  The work of M.G. has been financed by the Polish National Science Centre
  under the decision number DEC-2011/01/B/ST2/05932.
\end{ack}


\end{document}